\documentclass[prd,preprint,superscriptaddress,amsmath,amssymb,nofootinbib]{revtex4}
\usepackage{graphicx}
\usepackage{dcolumn}
\usepackage{bm}
\usepackage{amsmath}
\usepackage{epsfig}    
\usepackage{color}
\usepackage{slashed}
\usepackage{hhline}
\def\be{\begin{equation}}
\def\ee{\end{equation}}
\newcommand{\bea}{\begin{eqnarray}}
\newcommand{\eea}{\end{eqnarray}}
\newcommand{\nn}{\nonumber}

\numberwithin{equation}{section}

\begin{document}

\preprint{
\begin{minipage}{5cm}
\small
\flushright
EPHOU-25-001
\end{minipage}}

\title{ Zee-Babu model in a non-holomorphic modular  $A_4$ symmetry \\
and modular stabilization}

\author{Tatsuo Kobayashi}
\email{kobayashi@particle.sci.hokudai.ac.jp}
\affiliation{Department of Physics, Hokkaido University, Sapporo 060-0810, Japan}

\author{Hiroshi Okada}
\email{hiroshi3okada@htu.edu.cn}
\affiliation{Department of Physics, Henan Normal University, Xinxiang 453007, China}

\author{Yuta Orikasa}
\email{Yuta.Orikasa@utef.cvut.cz}
\affiliation{Institute of Experimental and Applied Physics, 
Czech Technical University in Prague, 
Husova 240/5, 110 00 Prague 1, Czech Republic}

\date{\today}

\begin{abstract}
We study a Zee-Babu neutrino model in a non-holomorphic modular $A_4$ symmetry,
and 
we construct a model so that there are minimum free parameters (two complex parameters for neutrino sector).
We find only the normal hierarchy is allowed.
Moreover, the allowed region to satisfy the neutrino oscillation data is localized at nearby $\tau=\omega$. The small absolute deviation plays a crucial role in fitting two mixings of $s^2_{23}$ and $s^2_{12}$.
In addition, we obtain several predictions on Majorana and Dirac CP phases, and neutrinoless double beta decay as shown in our chi square numerical analysis.  
We also discuss modulus stabilization within the framework of non-supersymmetric models.
In the end, we compute the expansion of modular forms at nearby $\tau=\omega$ in the Appendix 
so that one can apply them for a model and understand its analytical structure.
\end{abstract}

\maketitle

\section{Introduction}
A radiative seesaw model is one of the promising candidates to understand the tiny active 
neutrino masses in a low energy theory that reaches at current experiments such as CERN large hadron collider.
In particular, Zee-Babu(ZB) neutrino model provides the neutrino mass matrix at two-loop level 
as a leading contribution only by introducing a singly-charged boson $S^-$ and a doubly-charged boson $k^{++}$, and
it does not require any additional symmetries to forbid  tree level or one-loop level neutrino masses~\cite{Zee:1985id, Babu:1988ki}.
Moreover, the ZB model has two remarkable features.
The first one is that the lightest neutrino mass eigenvalue is zero because of the rank-two neutrino mass matrix that arises from anti-symmetric Yukawa $f$; $\overline {L_L}_i f_{ij} {L^c_L}_j S^+$. It implies that the two non-vanishing neutrino mass eigenvalues
are uniquely determined by inputting two observables, that is, solar and atmospheric mass square differences.
 {
The second one is that the loop function does not depend on the structure of neutrino mass matrix.
Therefore,  the number of free parameter, arising from the loop function, never increases.
This is because the charged-lepton masses, which run in the neutrino loop, can be negligible compared to the masses of $S^-$ and $k^{++}$.
} 

In this paper, we apply a non-holomorphic modular $A_4$ flavor symmetry for the ZB model
and investigate its nature of the lepton sector.
{For the last several years, modular flavor symmetric models have been being studied intensively \cite{Feruglio:2017spp}.
In those models, Yukawa couplings and masses are functions of the modulus $\tau$ and written by holomorphic modular forms of $S_3$, $A_4$, $S_4$, $A_5$ and others \cite{Feruglio:2017spp,Kobayashi:2018vbk,Penedo:2018nmg,Novichkov:2018nkm}.
Phenomenologically interesting results have been obtained within the framework of supersymmetric models.
(See for reviews refs.~\cite{Kobayashi:2023zzc,Ding:2023htn}.)}
The non-holomorphic modular symmetry was recently formulated by Qu and Ding~\cite{Qu:2024rns}, which can be applicable to non-supersymmetric theories without any potential difficulties. 
{That has extended the possibilities for model building including non-supersymmetric models with Yukawa couplings and masses of positive and negative weights.
Richer flavor structures can be realized by use of non-holomorphic modular forms.\footnote{When we start with supersymmetric models, supersymmetry breaking effects may affect the flavor structure at low-energy scale \cite{Kikuchi:2022pkd}.}}
After ref.~\cite{Qu:2024rns}, several ideas have been arisen in refs.~\cite{Nomura:2024atp, Ding:2024inn, Li:2024svh, Nomura:2024nwh, Nomura:2024vzw, Okada:2025jjo}. The non-holomorphic groups provide negative modular forms which do not have in holomorphic one.~\footnote{Note that the non-holomorphic modular symmetries include nature of holomorphic one.}
Thus, one expects different kinds of predictions would be obtained.
{In addition, the number of free parameters is less than the case of a holomorphic Zee-Babu model~\cite{Okada:2021aoi}
.~\footnote{This model favors a large Im[$\tau$], and both hierarchies satisfy the experimental results due to enough parameters.}}
In fact, we discover that our scenario has sharp predictions on mixing angles and phases only at nearby $\tau=\omega$, where $\omega=e^{2\pi i/3}$.
Here, we perform chi square numerical analysis and compare with predictions at $\tau=\omega$. Then, we find the small deviation crucially contributes to two reliable observables $s^2_{23}$ and $s^2_{12}$.
{We also discuss the modulus stabilization at nearby $\tau=\omega$ within the framework of non-supersymmetric models.}

This paper is organized as follows.
{In Sec.~\ref{sec:modular-forms}, we give a brief review on the modular symmetry and modular forms.}
In Sec.~\ref{sec:model}, we {explain} our model, constructing the valid charged-lepton Yukawa sector,  Higgs potential, and active neutrino sector. In the neutrino sector, we show several relations coming from features of the Zee-Babu model and modular $A_4$ symmetry. Then, we perform the $\Delta \chi^2$ analysis to get our results, comparing the result of $\tau=\omega$.
{In Sec.~\ref{sec:stabilization}, we discuss the modulus stabilization at nearby $\tau=\omega$ in non-supersymmetric models.}
We have conclusions and discussion in Sec.~\ref{sec:conclusion}.
{Appendix A shows $A_4$ modular forms, which we use.
In Appendix B, we show helpful formalisms which are expanded at nearby $\tau=\omega$.}
 {Appendix C shows the relationship between the values of $\Delta \chi^2$ and the confidence level. 
{Appendix D shows the result of neutrino mass matrix in case of $\tau=\omega$.}
 }

\section{Modular symmetry and Modular forms}
\label{sec:modular-forms}

We start with the homogeneous modular group $\Gamma=SL(2,\mathbb{Z})$, which is the special linear group of $2\times 2$ matrices with integer entries, 
\begin{align}
\gamma=
    \begin{pmatrix}
        a & b \\
        c & d
    \end{pmatrix},
\end{align}
where $ad-bc=1$.
This group is generated by the following two generators:
\begin{align}
    S=
    \begin{pmatrix}
        0 & 1 \\
        -1 & 0
    \end{pmatrix}, \qquad
    T=
    \begin{pmatrix}
        1 & 1 \\
        0 & 1
    \end{pmatrix}.
\end{align}
They satisfy the following algebraic relations:
\begin{align}
    S^2=-\mathbf{1}, \qquad (ST)^3=\mathbf{1},
\end{align}
where $\mathbf{1}$ is the identity.
The modular group includes the principal congruence subgroup of the level $N$, $\Gamma(N)$,  which is defined by 
\begin{align}
    \Gamma(N)=\left\{
    \begin{pmatrix}
        a & b \\
        c & d
    \end{pmatrix} 
    \in \Gamma, \quad
       \begin{pmatrix}
        a & b \\
        c & d
    \end{pmatrix} =
       \begin{pmatrix}
        1 & 0 \\
        0 & 1
    \end{pmatrix} 
    ~~~({\rm mod}~N)
    \right\}.
\end{align}
The quotients $\Gamma_N=\Gamma/\pm \Gamma(N)$ are isomorphic to $S_3$, $A_4$, $S_4$, and $A_5$ \cite{deAdelhartToorop:2011re}.\footnote{These non-Abelian groups have been used to explain the flavor structures of quarks and leptons \cite{Altarelli:2010gt,Ishimori:2010au,Hernandez:2012ra,King:2013eh,King:2014nza,Petcov:2017ggy,Kobayashi:2022moq}.}
In our model, we use $\Gamma_3 \simeq A_4$.

The modulus $\tau$ transforms 
\begin{align}
  \tau \to  \gamma \tau = \frac{a\tau + b}{c\tau +d},
\end{align}
under the modular symmetry.
Note that the point $\tau = \omega$ is the fixed point of the $(ST)$ transformation, where $\mathbb{Z}_3$ symmetry remains.
Also, the point $\tau=i$ is the fixed point of $S$.
Moreover, the $T$ symmetry remains in the limit 
${\rm Im}\tau \to \infty$.

The modular forms of weight $k$ at level $N$ transforms
\begin{align}
    Y^{(k)}_{\bf r}(\tau) \to  Y^{(k)}_{\bf r}(\gamma\tau)=(c\tau + d)^k \rho_{\bf r}(\gamma)Y^{(k)}_{\bf r}(\tau) ,
\end{align}
where $\rho_{\bf r}(\gamma)$ is a unitary matrix corresponding to the representation ${\bf r}$ of $\Gamma_N$.
For $k>0$, modular forms are holomorphic.
One can define non-holomorphic modular forms for $k \le 0$, i.e. polyharmonic Maa\ss~forms.
They satisfy the following Laplacian condition \cite{Qu:2024rns}:
\begin{align}
    \left( -4y^2 \frac{\partial}{\partial \tau} \frac{\partial}{\partial \bar \tau} +2iky \frac{\partial}{\partial \bar \tau} \right)Y^{(k)}_{\bf r}(\tau)=0,
\end{align}
where $\tau = x+iy$, 
and the proper growth condition,
\begin{align}
   Y^{(k)}_{\bf r}(\tau) ={\cal O}(y^\alpha),
\end{align}
as $y \to \infty$ for some $\alpha$.
The  polyharmonic Maa\ss~forms are non-holomorphic.

Matter fields $\phi$ of the weight $k$ and the representation ${\bf r}$ also transform as 
\begin{align}
    \phi \to (c\tau + d)^k\rho_{\bf r}(\gamma) \phi,
\end{align}
under the modular symmetry.
{The modular symmetry requires that each term in Lagrangian such as Yukawa coupling terms and the Higgs potential should have vanishing modular weight and be $\Gamma_N$ invariant.  }

We study a model with the $\Gamma_3 \simeq A_4$ symmetry.
Some modular forms of the level $N=3$ 
are shown in Appendix {A.}

\begin{table}[t!]
\begin{tabular}{|c||c|c||c|c|c|}\hline\hline  
& ~$\overline{ L_L}$~ & ~$ \ell_R$~ & ~$ H$ ~&~ {$S^-$} ~&~ {$k^{++}$}~  \\\hline\hline 
$SU(2)_L$   & $\bm{2}$  & $\bm{1}$  & $\bm{2}$  & $\bm{1}$& $\bm{1}$     \\\hline 
$U(1)_Y$    & $\frac12$  & $-1$ & $\frac12$ & {$-1$} & {$+2$}    \\\hline
$A_4$   & $\bm{3}$  & $ \bm{3} $  & $\bm{1}$ & $\bm{1}$& $\bm{1}$         \\\hline 
$-k_I$    & $-1$  & $+1$ & $0$ & $+2$ & $0$      \\\hline
\end{tabular}
\caption{Charge assignments of the SM leptons $\overline{ L_L}$ and $\ell_R$ and $\Delta$
under $SU(2)_L\otimes U(1)_Y \otimes A_4$ where $-k_I$ is the number of modular weight. Note here that all the bosons are assigned to be trivial singlet ${\bf1}$.
}\label{tab:1}
\end{table}

 \section{Model setup}
 \label{sec:model}
Here, we {explain} our setup based on the ZB model with a non-holomorphic modular $A_4$ symmetry.
Left-handed leptons, which are denoted by $\overline{L_L}\equiv[\bar\nu_L,\bar\ell_L]^T$,
are assigned to $A_4$ triplet with $-1$ modular weight. The right-handed charged-leptons, $\ell_R$,
are assigned to  $A_4$ triplet with $+1$ modular weight.
The singly-charged $S^-$ and doubly-charged bosons $k^{++}$ respectively have $+2$ and $0$ modular weights.
%
Note here that all the bosons are assigned to be trivial singlet ${\bf1}$.
The standard model Higgs is denoted by
{
$H\equiv[w^+,(v+h+iz)/\sqrt2]^T$ where $w^+$ and $z$ respectively give the gauge masses of $W^+$ and $Z$ in the standard model after the spontaneous symmetry breaking. $v\simeq246$ GeV is the vacuum expectation value.
}
The field contents and their assignments are summarized in Tab.~\ref{tab:1}.

\subsection{Charged-lepton mass matrix}
Under these symmetries, we write the renormalizability relevant Lagrangian for charged-leptons is found as follows:
\begin{align} 
&a_e   (\overline{L_{L_e}} e_R + \overline{L_{L_\mu}} \tau_R + \overline{L_{L_\tau}} \mu_R) H
+b_e\left[ y_1( 2\overline{L_{L_e}}  e_R -\overline{L_{L_\mu}} \tau_R -\overline{L_{L_\tau}} \mu_R) \right.\nn\\
& \left. + y_2( 2\overline{L_{L_\mu}}  \mu_R -\overline{L_{L_e}} \tau_R -\overline{L_{L_\tau}} e_R)
 + y_3( 2\overline{L_{L_\tau}}  \tau_R -\overline{L_{L_e}} \mu_R -\overline{L_{L_\mu}} e_R)\right] H \nn\\
&
+c_e\left[ y_1( \overline{L_{L_\tau}} \mu_R -\overline{L_{L_\mu}} \tau_R) + y_2( \overline{L_{L_\tau}} e_R -\overline{L_{L_e}} \tau_R)
 + y_3( \overline{L_{L_e}} \mu_R -\overline{L_{L_\mu}} e_R)\right] H
+  {\rm h.c.} ,
\label{eq:cgdlptn}
\end{align}
where $Y^{(0)}_3\equiv[y_1,y_2,y_3]^T$~\cite{Qu:2024rns}.
(See Appendix A.)
After the spontaneous symmetry breaking of $H$,
the charged-lepton mass matrix $M_\ell$ is given by
\begin{align}
M_\ell  = \frac{v}{\sqrt2}
\left[
a_e
\begin{pmatrix}
1 & 0 & 0 \\ 
 0 & 0 &1 \\ 
0 & 1 & 0 \\ 
\end{pmatrix}
+ 
b_e
\begin{pmatrix}
2 y_1 & -y_3 & -y_2 \\ 
 -y_3 &2 y_2 & -y_1 \\ 
-y_2 & -y_1 & 2y_3 \\ 
\end{pmatrix}
+ 
c_e
\begin{pmatrix}
 0 & y_3 & -y_2 \\ 
 -y_3 &0 & y_1 \\ 
y_2 & -y_1 & 0 \\ 
\end{pmatrix}
\right]
 \label{massmat},
\end{align}
where  $a_e, b_e, c_e$, which are real parameters {for simplicity}, 
are determined in order to fix the mass eigenvalues of charged-leptons.
Inputting the experimental masses of charged-leptons and a value of $\tau$, these free parameters are numerically solved by the following three relations:
{
\begin{align}
&{\rm Tr}[M_\ell M_\ell^\dag] = m_e^2 + m_\mu^2 + m_\tau^2,\\
&{\rm Det}[M_\ell M_\ell^\dag] = m_e^2  m_\mu^2  m_\tau^2,\\
&({\rm Tr}[M_\ell M_\ell^\dag])^2 -{\rm Tr}[(M_\ell M_\ell^\dag)^2] =2( m_e^2  m_\mu^2 + m_\mu^2  m_\tau^2+ m_e^2  m_\tau^2 ).
\end{align}
Here, $M_\ell$ is diagonalized by bi-unitary mixing matrices as $D_\ell\equiv{\rm diag}(m_e,m_\mu,m_\tau)=V^\dag_{eL} M_\ell V_{eR}$.
$m_e,m_\mu,m_\tau$ are mass eigenvalues for the charged-leptons and precisely measured by experiments. We refer to Particle Data Group~\cite{ParticleDataGroup:2022pth} for these experimental values.}

\subsection{Higgs potential}
Renormalizable Higgs potential is given by 
\begin{align}
  {\cal V} &= -\mu^2_H |H|^2-\mu^2_S |S^-|^2-\mu^2_k |k^{++}|^2 + (\mu S^- S^- k^{++}   + {\rm h.c.})\\
  &+\lambda_H |H|^4 + \lambda_S |S^-|^4 +\lambda_k |k^{++}|^4 + \lambda_{HS}|H|^2|S^-|^2+ \lambda_{Hk}|H|^2|k^{++}|^2
  + \lambda_{Sk}|S^-|^2|k^{++}|^2 
  ,\label{Eq:pot}
\end{align}
where $ \mu_S^2 \equiv \mu_{S_0}^2 (\tau-\bar \tau)^{2}$,  $ \mu\equiv \mu_0 Y^{(-4)}_1$, $ \lambda_S \equiv \lambda_{S_0} |(\tau-\bar \tau)^{2}|^2$,
$\lambda_{HS} \equiv \lambda_{HS_0} (\tau-\bar \tau)^{2}$,
$\lambda_{Sk} \equiv \lambda_{Sk_0} (\tau-\bar \tau)^{2}$ in order to be invariant under the modular symmetry. 
After spontaneous symmetry breaking by $H$, physical Higgs masses are found as
\begin{align}
m^2_H &= \lambda_H v^2,\\
m^2_s &= -\mu^2_S + \frac{\lambda_{HS}}{2} v^2,\\
m^2_k &= -\mu^2_k + \frac{\lambda_{Hk}}{2} v^2.
\end{align}

\subsection{Active neutrino mass matrix}
\label{neut}
The valid Lagrangian for the neutrino sector is found as follows:
\begin{align}
-{\cal L}_\nu &= a_s\left[y_1(\overline{L_{L_\mu}}\cdot {L^c_{L_\tau}} -\overline{L_{L_\tau}}\cdot {L^c_{L_\mu}}) \right.\nn\\
&\left.
+y_2(\overline{L_{L_\tau}}\cdot {L^c_{L_e}} -\overline{L_{L_e}}\cdot {L^c_{L_\tau}})
+y_3(\overline{L_{L_e}}\cdot {L^c_{L_\mu}} -\overline{L_{L_\mu}}\cdot {L^c_{L_e}})\right]S^-\nn\\
&  +
a_k\left[ y'_1 (2 \overline{e^c_R} e_R -\overline{\mu^c_R} \tau_R - \overline{\tau^c_R} \mu_R) \right.\nn\\
&\left.
+y'_2 (2 \overline{\mu^c_R} \mu_R -\overline{e^c_R} \tau_R - \overline{\tau^c_R} e_R) 
+y'_3 (2 \overline{\tau^c_R} \tau_R -\overline{e^c_R} \mu_R - \overline{\mu^c_R} e_R)  \right]k^{++}\\
& + b_k \left[\overline{e^c_R} e_R + \overline{\mu^c_R} \tau_R + \overline{\mu^c_R} \tau_R\right]  k^{++} +{\rm h.c.}
\label{eq:neut}\\
&=
a_s\left[
\overline{\nu_L} f V^*_{eL} \ell^c_L +\overline{\ell_L} V^\dag_{eL} f^T \nu^c_L
\right] S^-
+
a_k \overline{\ell^c_R} V^T_{eR} g V_{eR} \ell_R k^{++} +{\rm h.c.},
\end{align}
where {$b_k$ includes $Y^{(-2)}_1$}, we define $\ell_{L/R}\equiv [e,\mu,\tau]^T_{L,R}$ which are mass eigenstates of the charged-leptons in the last line, 
$\cdot \equiv i\tau_2$, where $\tau_2$ is second Pauli matrix,  and $Y^{(-2)}_3\equiv[y'_1,y'_2,y'_3]^T$~\cite{Qu:2024rns}. (See Appendix A.)
Then, the neutrino mass matrix is given at two-loop level as follows:
\begin{align}
  (m_{\nu})_{ij}
 &\simeq \frac{\mu^*}{64\pi^2} \frac{a_s^2 a_k}{m_s^2}I(r) 
 \left[f V^*_{eL} D^*_\ell V^T_{eR} g V_{eR} D^\dag_\ell V^\dag_{eL} f^T\right]_{ij}
 \equiv  \kappa 
\left[f M^*_\ell g M^\dag_\ell f^T\right]_{ij}
 ~,
\label{massmatrix2}\\
&I(r) = -\int_0^1 dx \int_0^{1-x} dy \frac{1-y}{x+(r-1)y+y^2} \ln\left[\frac{y(1-y)}{x+r y}\right],\\
f&= \begin{pmatrix}
0  &  y_3 & -y_2 \\ 
-y_3 & 0 & y_1 \\ 
y_2 & -y_1 & 0 \\ 
\end{pmatrix}
,\quad
g= \begin{pmatrix}
2 y'_1  & -y'_3 &-y'_2 \\ 
-y'_3 & 2y'_2 & -y'_1 \\ 
-y'_2 & -y'_1 & 2y'_3 \\ 
\end{pmatrix}
+\tilde b_k
\begin{pmatrix}
1  & 0 & 0 \\ 
0 & 0 & 1 \\ 
0 & 1 & 0 \\ 
\end{pmatrix}
,
\end{align}
where $\tilde b_k\equiv b_k/a_k$, $r\equiv m_k^2/m_s^2$. Since the loop function $I(r)$ does not depend on the masses of charged-leptons in the limit of $\frac{m_{e,\mu,\tau}}{m_{s,k}}<<1$, one can redefine $m_\nu\equiv \kappa \tilde m_\nu $.
{
Here, $\kappa$ is an overall parameter of $m_\nu$ and defined by
\begin{align}
\kappa\equiv  \frac{\mu^*}{64\pi^2} \frac{a_s^2 a_k}{m_s^2}I(r) .
\end{align}
}
 Since the coupling $f$ is an anti-symmetric matrix that is rank two, $ \tilde m_\nu $ has one vanishing mass eigenvalue.
 Therefore, we find
 \begin{align}
&( {\rm NH}):\quad V^T_\nu \tilde m_\nu V_\nu ={\rm diag}(0,\tilde D_{\nu_2},\tilde D_{\nu_3}),\\
&( {\rm IH}):\quad V^T_\nu \tilde m_\nu V_\nu ={\rm diag}(\tilde D_{\nu_1},\tilde D_{\nu_2},0).
 \end{align}

 In this case, one can write $\kappa^2$ in terms of either of two mass squared differences $\Delta m^2_{\rm atm}$ and $\Delta m^2_{\rm sol}$,
and mass eigenvalues of $\tilde m_\nu$ as follows:
\begin{align}
({\rm NH}):\  \kappa^2= \frac{|\Delta m_{\rm atm}^2|}{\tilde D_{\nu_3}^2},
\quad
({\rm IH}):\  \kappa^2= \frac{|\Delta m_{\rm atm}^2|}{\tilde D_{\nu_2}^2},\label{eq:kappa}
 \end{align}
 where we choose $\Delta m^2_{\rm atm}$ as an input observable.
Applying Eq.~(\ref{eq:kappa}), $\Delta m^2_{\rm sol}$ is given by
\begin{align}
& ({\rm NH}):\   \Delta m_{\rm sol}^2= {\kappa^2} {\tilde D_{\nu_2}^2}
=\frac{\tilde D_{\nu_2}^2}{\tilde D_{\nu_3}^2} |\Delta m_{\rm atm}^2|,\\
& ({\rm IH}):\   \Delta m_{\rm sol}^2= {\kappa^2}({\tilde D_{\nu_2}^2-\tilde D_{\nu_1}^2})
=\left(1-\frac{\tilde D_{\nu_1}^2}{\tilde D_{\nu_2}^2}\right) |\Delta m_{\rm atm}^2|.
\end{align}
Sum of neutrino masses $\sum D_{\nu_i}$ is found as follows:
\begin{align}
& ({\rm NH}):\   \sum D_{\nu_i} \sim \sqrt{\Delta m_{\rm atm}^2}\sim60\ {\rm GeV},\label{eq:mrel}\\
& ({\rm IH}):\   \sum D_{\nu_i} \sim 2 \sqrt{\Delta m_{\rm atm}^2}\sim120\ {\rm GeV}.
\end{align}
{A recent result of  DESI and CMB data combination, which provides  the upper bound~$\sum D_{\nu}\le$ 72 meV~\cite{DESI:2024mwx},
favors only the case of NH.
On the other hand, 
The minimal standard cosmological model with CMB data~\cite{Planck:2018vyg}; $\sum D_{\nu} \le 120 \, {\rm meV}$,
allows both cases of these hierarchies.
}
%

The observed mixing matrix is defined by $U=V_{e_L}^\dag V_\nu$~\cite{Maki:1962mu}
and the mixing angles are given in terms of the component of $U$ as follows:
\begin{align}
\sin^2\theta_{13}=|U_{e3}|^2,\quad 
\sin^2\theta_{23}=\frac{|U_{\mu3}|^2}{1-|U_{e3}|^2},\quad 
\sin^2\theta_{12}=\frac{|U_{e2}|^2}{1-|U_{e3}|^2}.
\end{align}
Also, we compute the Jarlskog invariant, $\delta_{CP}$ derived from PMNS matrix elements $U_{\alpha i}$:
\begin{equation}
J_{CP} = \text{Im} [U_{e1} U_{\mu 2} U_{e 2}^* U_{\mu 1}^*] = s_{23} c_{23} s_{12} c_{12} s_{13} c^2_{13} \sin \delta_{CP},
\end{equation}
where $s_{ij}\equiv \sin \theta_{ij}$ and $c_{ij}\equiv \cos \theta_{ij}$. 
The Majorana phases are found in terms of the following relations:
\begin{align}
& \text{Im}[U^*_{e1} U_{e2}] = c_{12} s_{12} c_{13}^2 \sin \left( \frac{\alpha_{21}}{2}\right), \
 \text{Re}[U^*_{e1} U_{e2}] = c_{12} s_{12} c_{13}^2 \cos \left( \frac{\alpha_{21}}{2} \right), 
\\
&
\text{Im}[U^*_{e1} U_{e3}] = c_{12} s_{13} c_{13} \sin \left( \frac{\alpha_{31}}{2} - \delta_{CP} \right), \
 \text{Re}[U^*_{e1} U_{e3}] = c_{12} s_{13} c_{13} \cos \left( \frac{\alpha_{31}}{2} - \delta_{CP} \right),  
\end{align}
and $\alpha = \alpha_{21} - \alpha_{31}$.
%
Using these mixing angles and phases, the neutrinoless double beta decay $\langle m_{ee}\rangle$ is described by 
\begin{align}
& ({\rm NH}):\ \langle m_{ee}\rangle=\kappa|\tilde D_{\nu_2} s^2_{12} c^2_{13}
e^{i\alpha}+\tilde D_{\nu_3} s^2_{13}e^{-2i\delta_{CP}}|,\\
& ({\rm IH}):\ \langle m_{ee}\rangle=\kappa|\tilde D_{\nu_1} c^2_{12} c^2_{13}+\tilde D_{\nu_2} s^2_{12} c^2_{13}
e^{i\alpha}|.
\end{align}
$\langle m_{ee}\rangle$ would be measured by future experiment of KamLAND-Zen~\cite{KamLAND-Zen:2024eml},
{and
its upper bounds are given by $ \langle m_{ee}\rangle <(28-122)$ meV at 90 \% confidence level.}

\subsection{Numerical analysis \label{sec:NA}}
We show our $\Delta \chi^2$ analysis, applying the neutrino oscillation data at 5$\sigma$ intervals in NuFit6.0~\cite{Esteban:2024eli} 
where Dirac CP phase is considered as an output parameter.
Therefore, we have five fitting observables: $ \Delta m_{\rm sol}^2,\Delta m_{\rm atm}^2,s_{12}, s_{13},s_{23}$. 
The masses of charged-leptons are referred to ref.~PDG\cite{ParticleDataGroup:2022pth}.
 Our input parameters are randomly selected within the following ranges:
 \begin{align}
 |\tilde b_k|\in [0, 100],\quad {\rm arg}[\tilde b_k]\in [-\pi, +\pi], \label{eq:paraspa}
 \end{align}
 where modulus $\tau$ is within the range of the fundamental region.
Since we do not have a solution within the range of $5\sigma$ in the case of IH, we shall now concentrate on discussing NH. 
{
Interestingly, we have solutions only at nearby $\tau=\omega$ with the range of $4\sigma - 5\sigma$ and minimum $\Delta \chi^2$ is about 16. Thus, all the plots are within this range.
{
We summarize allowed ranges of our parameters with minimum $\Delta\chi^2$ in Tab.~\ref{allowedrange}, where
 the value of atmospheric mass square difference runs over whole the experimental regions $\frac{\Delta m^2_{\rm atm}}{{[\rm eV^2}]} \times10^{3}=[2.463,2.606]$ at 3$\sigma$ range~\cite{Esteban:2024eli}.
One clearly finds that $s^2_{12}$ at $\tau\simeq \omega$ is deviated from experimental results at 3$\sigma$ interval. }

\begin{table}[tb]
	\centering
	\begin{tabular}{|c||c||c|} \hline 
		\rule[14pt]{0pt}{0pt}	  &  $\tau \simeq \omega$       &  
		$  {\rm Exp. (\le3\sigma)}$~\cite{Esteban:2024eli} \\ \hline 
		\rule[14pt]{0pt}{0pt}	
		${\rm Re}[\tau]$ &   $ [-0.496417,-0.495257]$     & $-$    \\ 
\rule[14pt]{0pt}{0pt}	
		${\rm Im}[\tau]$ &   $  [0.869247,0.872521]$        & $-$  \\ 
${\rm Abs}[\tilde b_k]$ &   $ [0.392093, 0.630404 ]$        & $-$   \\  
\rule[14pt]{0pt}{0pt}	
		$|{\rm Arg}[\tilde b_k]|$ &   $  [1.67832, 2.46345]$      & $-$  \\ \hline
		$\frac{\Delta m^2_{\rm sol}}{[\rm eV^2]}\times10^{5} $ & $ [7.07173,8.09637]$  & $ [6.92, 8.05]$ \\
		$s^2_{13}$ & $ [0.0209175, 0.023878]$	  & $ [0.02023, 0.02376]$ \\
		$s^2_{23}$ & $ [0.430074, 0.440551]$	  & $ [0.430, 0.596]$ \\
		$s^2_{12}$ & $ [0.35123, 0.377646]$	 & $ [0.275, 0.345]$\\
		$\delta_{CP}/$[deg] & $ [-35.245, -2.506]$	 & $ [-264, 62]$ \\\hline\hline
		$\Delta \chi^2_{\rm min}$ & $16.0067 $	   & $0.6$ \\
		\hline
	\end{tabular}
	\caption{Allowed regions for our input and observed parameters for NH. Here, the value of atmospheric mass square difference runs over whole the experimental regions $\frac{\Delta m^2_{\rm atm}}{{[\rm eV^2}]} \times10^{3}=[2.463,2.606]$ at 3$\sigma$ range~\cite{Esteban:2024eli}.
{Clearly, $s^2_{12}$ at $\tau\simeq \omega$ is deviated from experimental results at 3$\sigma$ interval. }
	}
	\label{allowedrange}
\end{table}

\begin{figure}[htbp]
  \includegraphics[width=77mm]{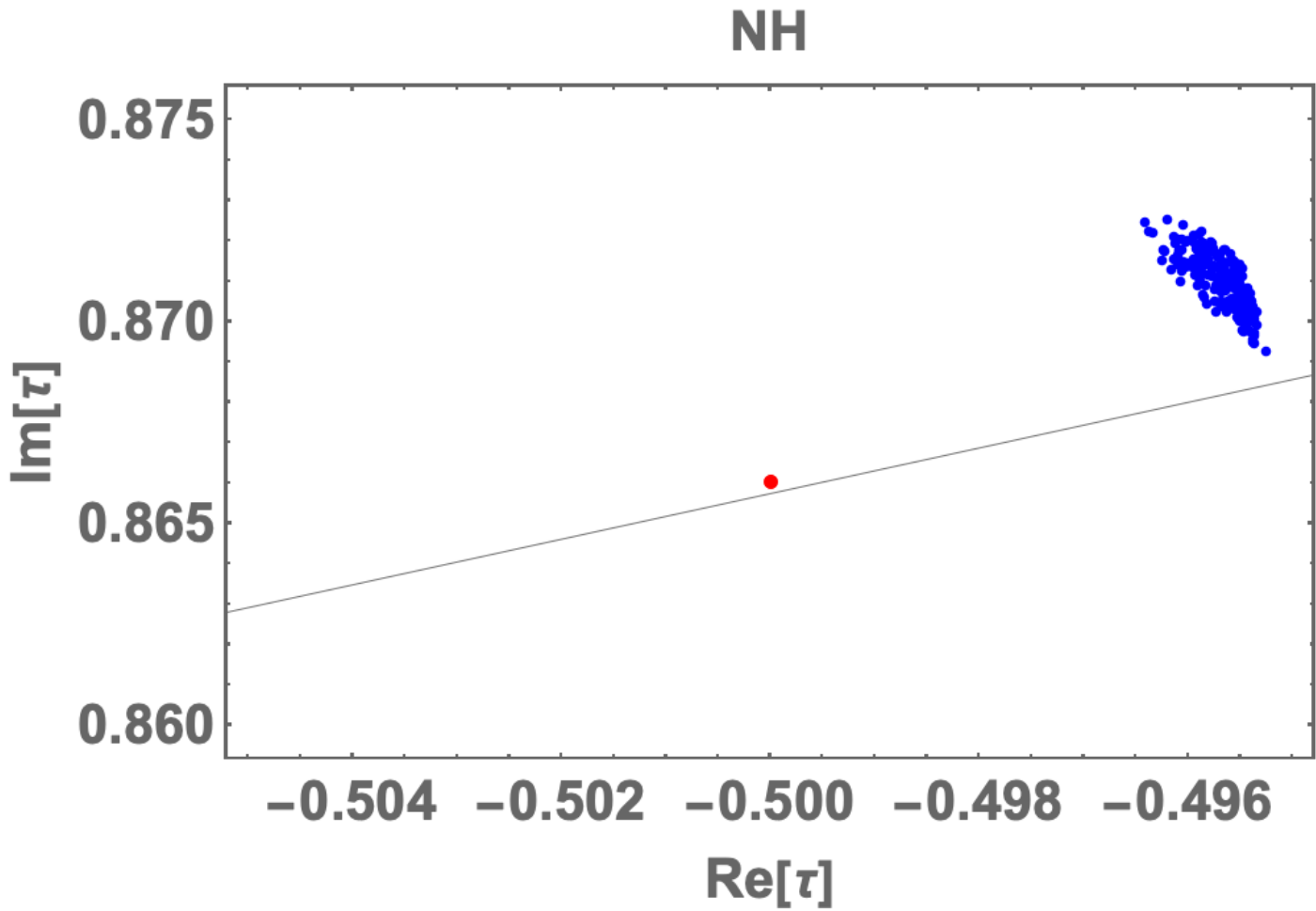} 
  \caption{Allowed region in terms of real $\tau$ and imaginary $\tau$.
    The blue points are localized at the region at nearby $\omega\equiv e^{2\pi i}$ where red point is $\tau=\omega$}
  \label{fig:tau_nh}
\end{figure}
Fig.~\ref{fig:tau_nh} shows the allowed region of $\tau$ within the fundamental region where the blue points are localized at the region nearby $\omega\equiv e^{2\pi i/3}$.
The red point represents $\tau=\omega$.
\footnote{
We show the result in case of  $\tau=\omega$ in Appendix D, in which we find that there are two massless of the neutrinos masses.
Therefore, it is totally ruled out by experiment. 
}
Interestingly, $\tau$ is very close to $\omega$.
The absolute value of deviation from $\tau=\omega$ is about 0.006.
We also show the allowed ranges of our input parameters $\tau$ and $\tilde b_k$ in Tab.~\ref{allowedrange} through our numerical analysis.
In case of $\tau\simeq \omega$, $|\tilde b_k|\approx[0.39, 0.63]$ and $|{\rm Arg}[b_k]|\approx[1.68, 2.46]$ are requested which is rather localized compared to our parameter space in Eq.~(\ref{eq:paraspa}). 


\begin{figure}[htbp]
  \includegraphics[width=77mm]{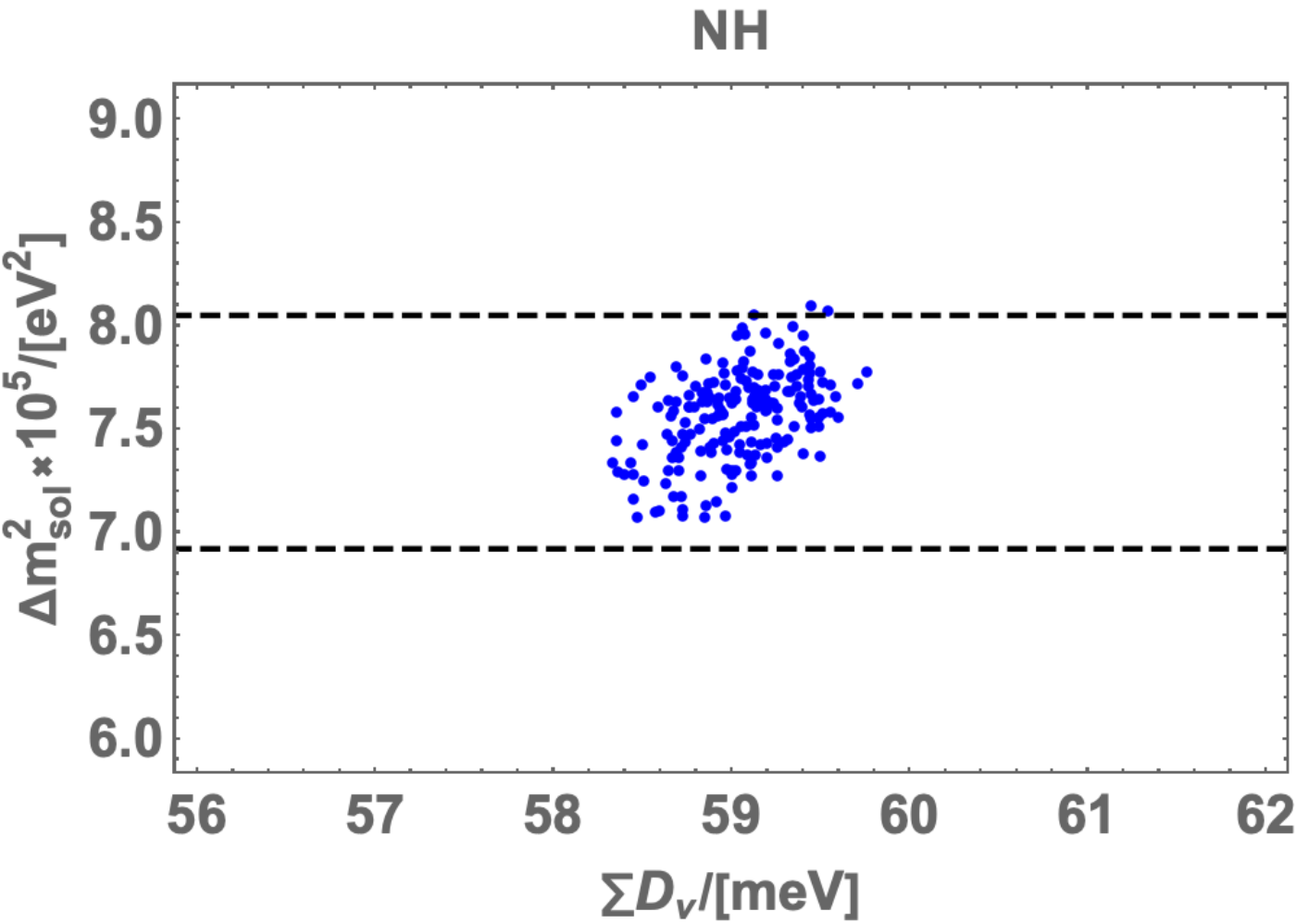}
  \includegraphics[width=77mm]{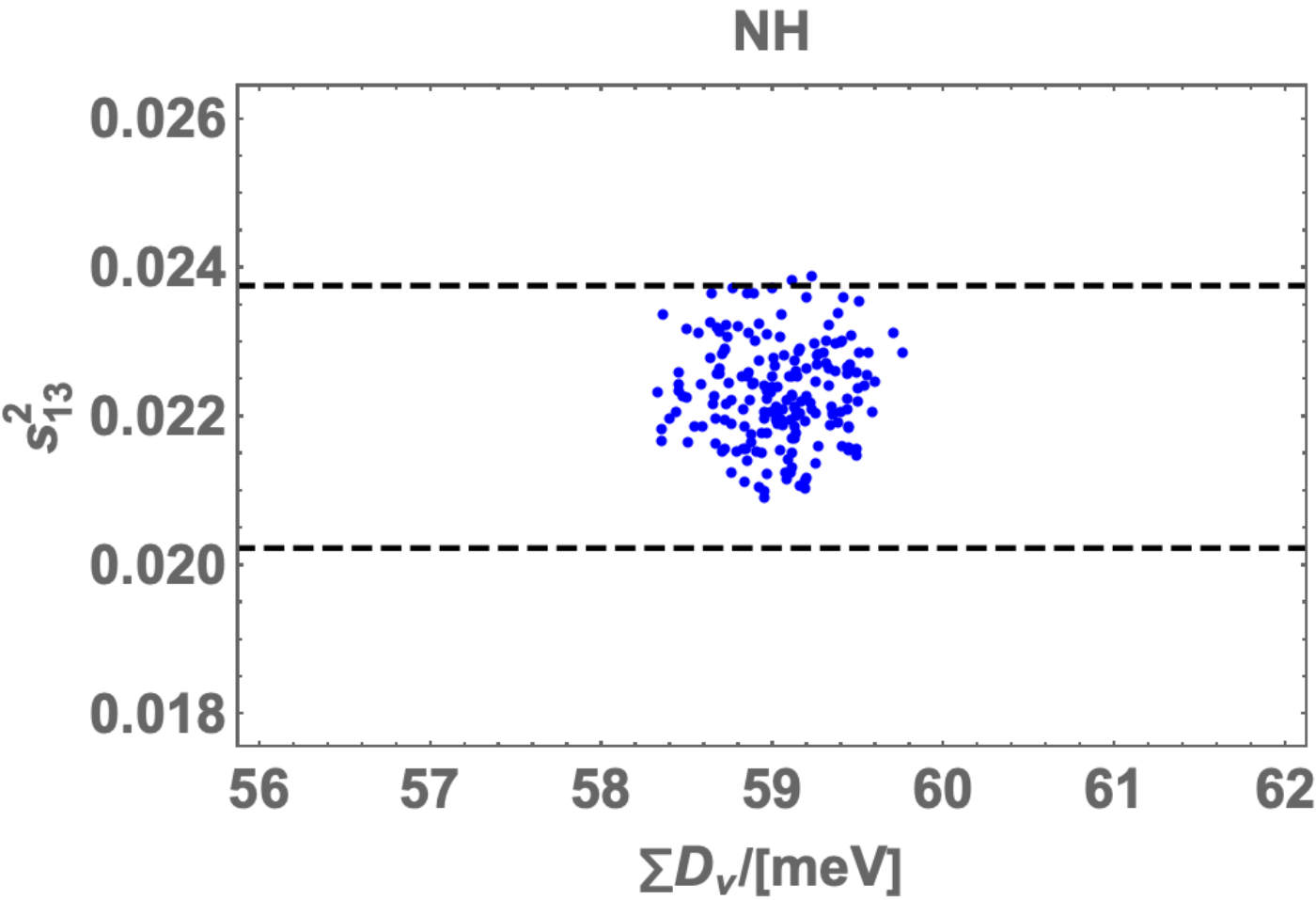}\\
  \includegraphics[width=77mm]{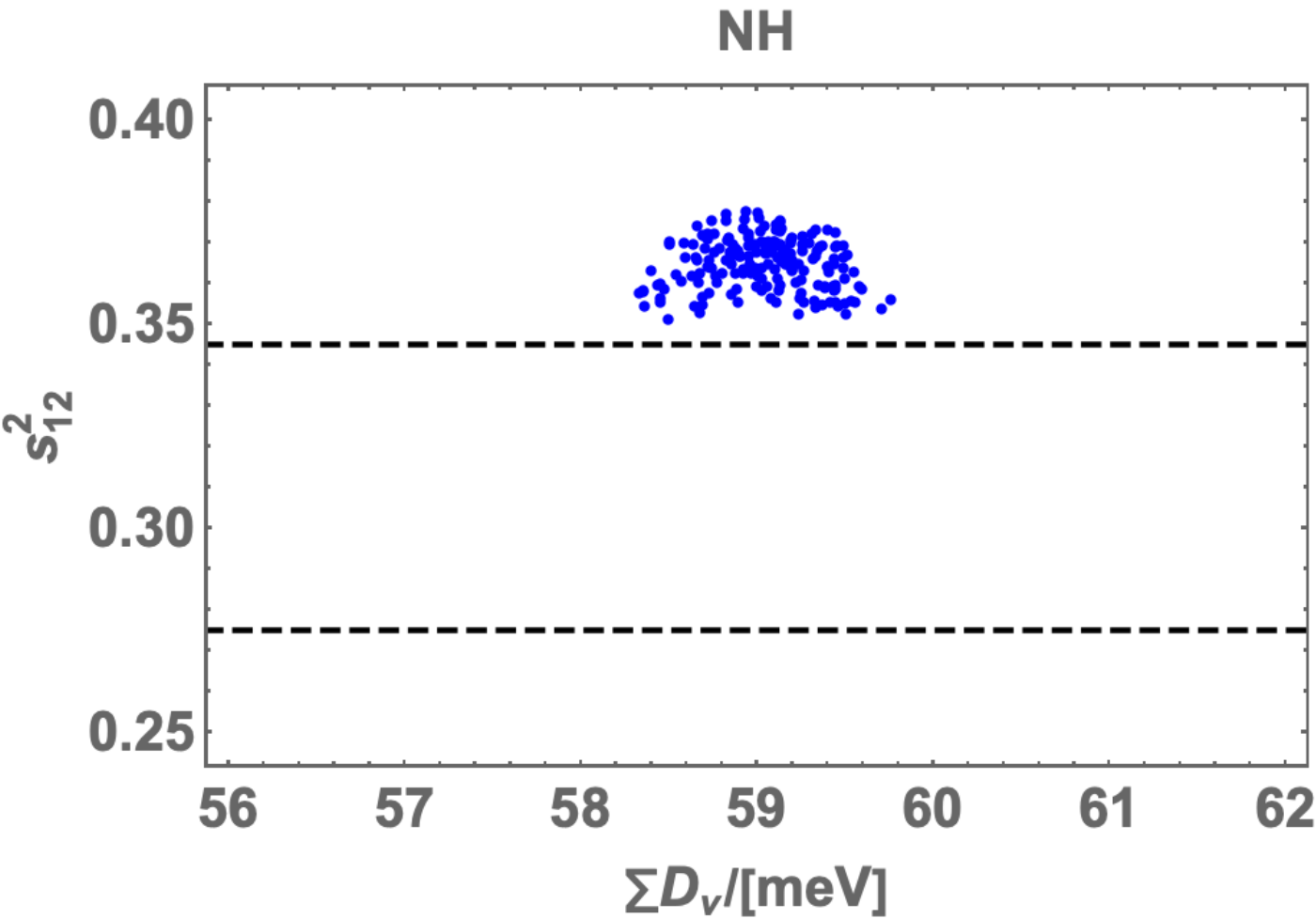}
  \includegraphics[width=77mm]{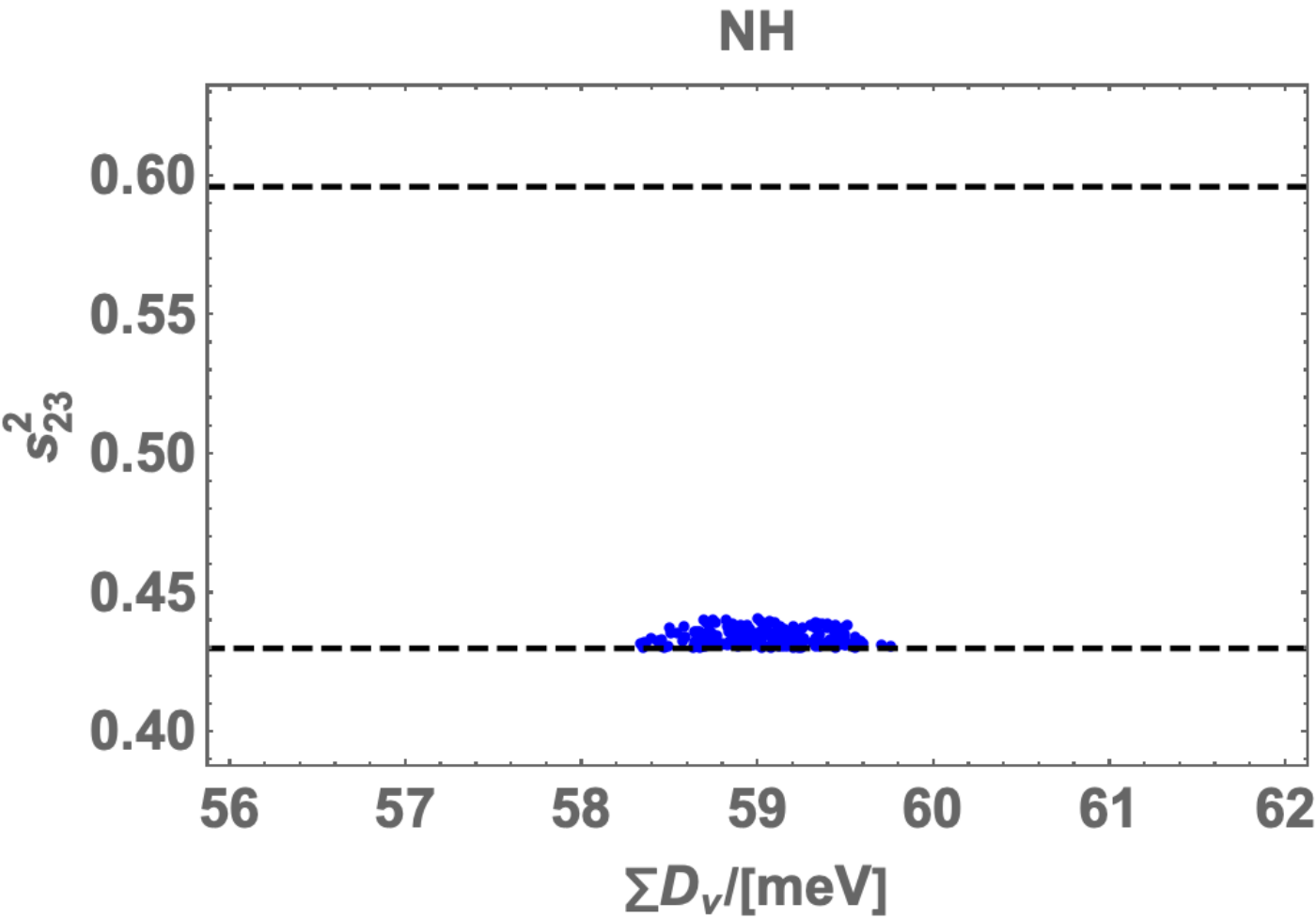}
  \caption{Blue points are our allowed regions of $\Delta m^2_{\rm sol}$(up-left), $s^2_{13}$(up-right), $s^2_{12}$(down-left), and $s^2_{23}$(down-right)  in terms of  $\sum D_{\nu}$.
Each dotted horizontal black line represents experimental upper(lower)-bound at 3$\sigma$ interval where these lines are given by two dimensional $\chi^2$ analysis. 
} \label{fig:reliable-obs}
\end{figure}
Fig.~\ref{fig:reliable-obs} shows our allowed regions (with blue points) of $\Delta m^2_{\rm sol}$(up-left), $s^2_{13}$(up-right), $s^2_{12}$(down-left), and $s^2_{23}$(down-right)  in terms of  $\sum D_{\nu}$.
Each dotted horizontal black line represents experimental upper(lower)-bound at 3$\sigma$ interval where these lines are given by two dimensional $\chi^2$ analysis.
%
The range of $\sum D_\nu$ is about [58-60] meV that is nothing but the direct result of Eq.~(\ref{eq:mrel}),
satisfying the upper bound on $\sum D_\nu\le 72$ eV from the combined result of DESI and CMB.
These figures suggest that $\Delta m^2_{\rm sol}$ and $s^2_{13}$ almost run within the whole range of experimental ranges. But $s^2_{23}$ is localized at nearby the lower experimental bound.
$s^2_{12}$ is slightly deviated from the upper experimental bound.
~\footnote{Our analysis is computed via five dimensional $\chi^2$ analysis and this deviation does not mean our model is ruled out by the experimental result(that is calculated by two dimensional analysis). }
In Tab.~\ref{allowedrange}, we show allowed ranges for these observables for $\tau\simeq\omega$ and experimental results at $3\sigma$ interval.
%

\begin{figure}[htbp]
  \includegraphics[width=77mm]{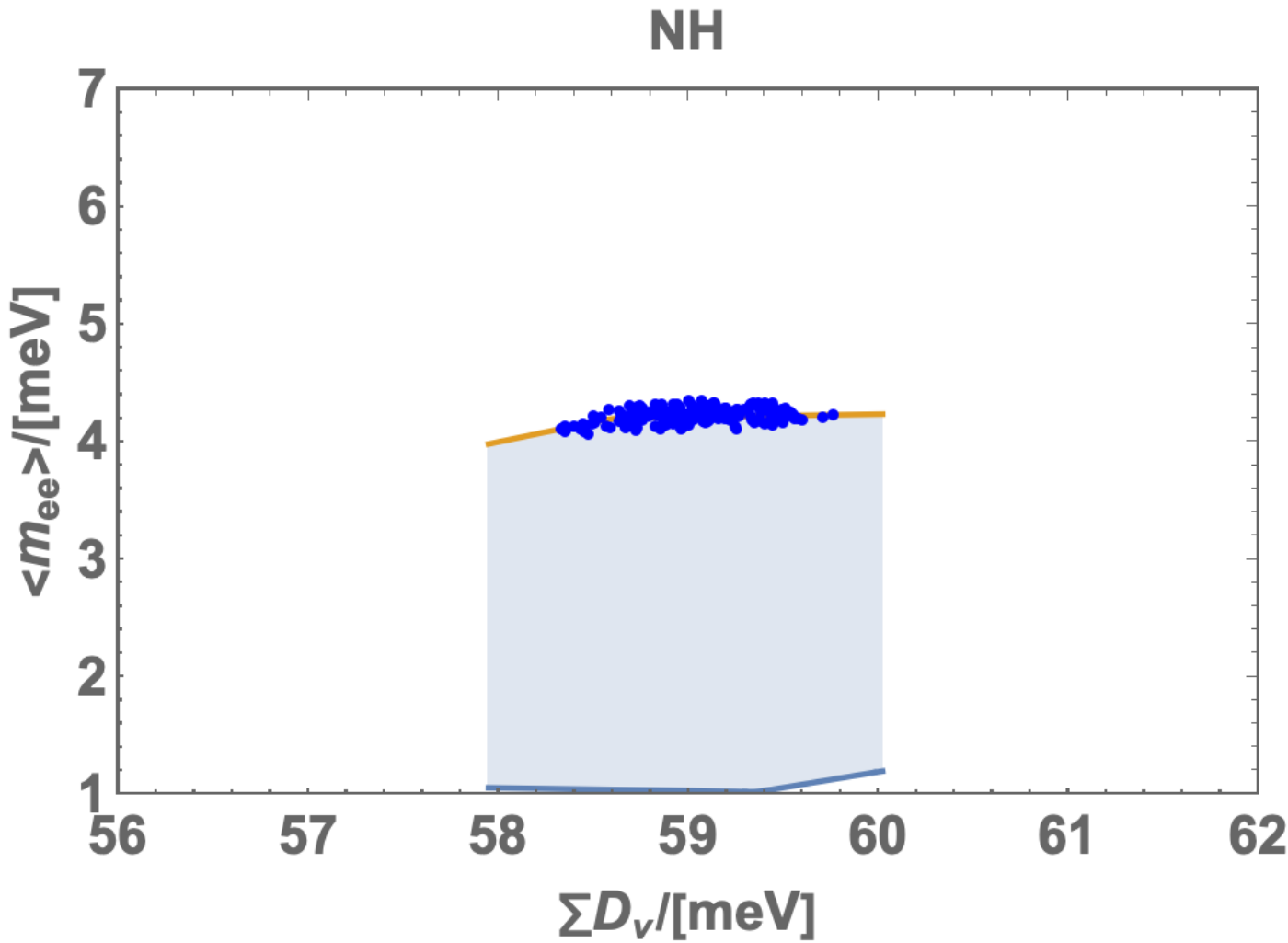}
  \caption{Allowed region for the neutrinoless double beta decay in terms of sum of neutrino masses.
 The legend of colors are the same as Fig.1.
 {The shaded regions are allowed by the current experimental data at 3 $\sigma$.}}
  \label{fig:masses_nh}
\end{figure}
Fig.~\ref{fig:masses_nh} shows allowed region  between $\langle m_{ee}\rangle$ and $\sum D_\nu$ in meV unit, where the legends of colors are the same as Fig.~1. 
%
The range of $\langle m_{ee}\rangle$ is about  [4-4.4] meV for $\tau\simeq\omega$ which is much below the current upper bound of KamLAND-Zen.
Here, the shaded regions are allowed by the current experimental data at 3 $\sigma$. 

\begin{figure}[htbp]
  \includegraphics[width=77mm]{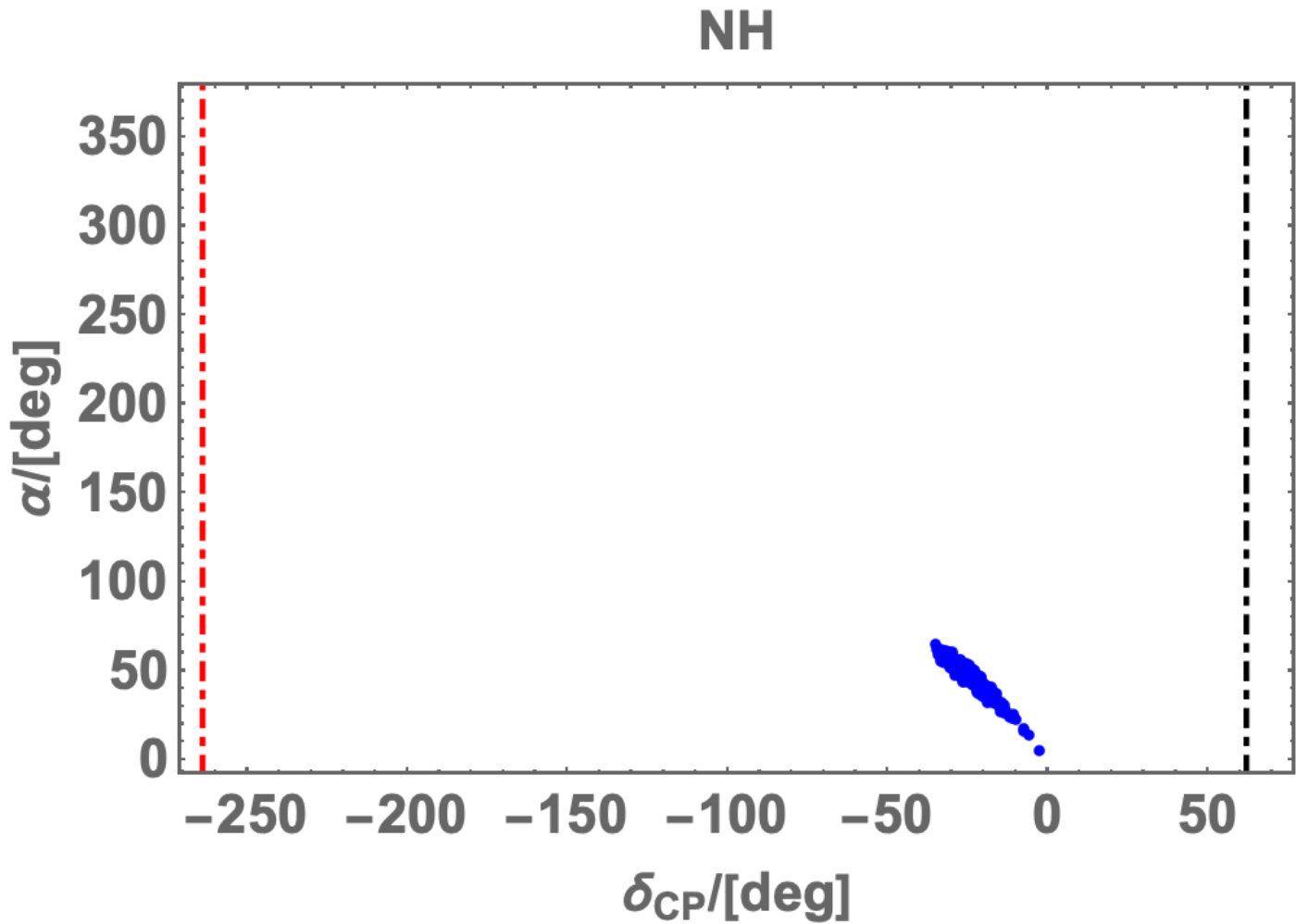}
  \caption{The scatter plots for the Majorana phase $\alpha$ in terms of Dirac CP phase $\delta_{\rm CP}$, where {the red dashed-dot vertical line at
-264 [deg] and the black one at 62 [deg] are respectively the lower and upper bound on Nufit 6.0~\cite{Esteban:2024eli}.}
The legend of colors are the same as Fig.1.}
  \label{fig:phase_nh}
\end{figure}
Fig.~\ref{fig:phase_nh} shows correlation between $\delta_{\rm CP}$ and $\alpha$, where the legend of colors are the same as Fig.1.
{The red dashed-dot vertical line at
-264 [deg] and the black one at 62 [deg] are respectively the lower and upper bound on Nufit 6.0~\cite{Esteban:2024eli} at 3 $\sigma$.
}
It reads that Majorana phase tends to be localized at the range of [0, 65] deg, while the range of Dirac CP phase is about [-35, -2.5] [deg] in case of $\tau\simeq\omega$.~\footnote{If we include the case of $\tau\simeq-\omega^2$, {which is equivalent to $\tau=\omega$, but out of the fundamental domain including $\tau=\omega$} our allowed region becomes to be symmetric.}
%

}

{
It would be worth mentioning the same sign charged-lepton decays of $k^{--}$, induced by the Yukawa coupling $g$, exhibit a characteristic pattern reflecting the structure of $g$.
In the following, we simply assume that $k^{--}$ decays into $\ell_a\ell_b$ with a branching fraction of 100 \%. Under this assumption, each of the branching ratio is approximately given by
\begin{align}
& {\rm BR}(k^{--}\to  e^-e^-)\approx
25 
\  \%,\ 
 {\rm BR}(k^{--}\to  \mu^-\mu^-)\approx 
 25 
 \  \%,\ 
 {\rm BR}(k^{--}\to  \tau^-\tau^-)\approx 
 15 
 \  \%, \nn\\
&  {\rm BR}(k^{--}\to e^-\mu^-) \approx 
15 
\  \%,\ 
  {\rm BR}(k^{--}\to e^-\tau^-) \approx 
  10 
  \  \%,\ 
  {\rm BR}(k^{--}\to \mu^-\tau^-) \approx 
  10 
  \  \% .
\end{align} 
The features of the branching ratios are rather universal; however, $k^{--}$ predominantly decays into $e^-e^-$ and $\mu^-\mu^-$ , which together account for approximately 50 \% of the total decay width.

}

\section{Modulus stabilizaiton}
\label{sec:stabilization}

As shown in the previous section, our model leads to realistic results at nearby $\tau=\omega$.
In this section, we discuss a possibility for the modulus stabilization to realize such a value.
In these years, modulus stabilization has been studied within the framework of supersymmetric modular flavor models \cite{Kobayashi:2019xvz,Kobayashi:2019uyt,Abe:2020vmv,Ishiguro:2020tmo,Novichkov:2022wvg,Knapp-Perez:2023nty,King:2023snq,Funakoshi:2024yxg,Abe:2024tox,Higaki:2024pql}.
Some of those scenarios lead to the modulus stabilization at $\tau = \omega$.
Here, we discuss a possibility for modulus stabilization in a non-supersymmetric model leading to $\tau = \omega$.

For our purpose, it is important to understand the behavior of modular forms around $\tau = \omega$.
The point $\tau = \omega$ is the fixed point of the $ST$ transformation, and the $\mathbb{Z}_3$ symmetry remains.
Under the $ST$ transformation, the modular form transforms as{
\begin{align}
 Y^{(k)}_{\bf r}(ST \tau,ST\bar \tau) = (-\tau -1)^k \rho_{\bf r}(ST) Y^{(k)}_{\bf r}(\tau,\bar \tau).
\end{align}
}
It is convenient to use the parameter $u$ \cite{Novichkov:2021evw},
\begin{align}
u \equiv \frac{\tau -\omega}{\tau -\omega^2}.
\end{align}
We also use its complex conjugate $\bar u$,
\begin{align}
\bar u \equiv \frac{\bar \tau -\omega^2}{\bar \tau -\omega},
\end{align}
when we discuss non-holomorphic modular forms.
{See Appendix B  for the behavior of modular forms around $\tau =\omega$.}
{For the purposes of the following discussion, we will focus on  $Y^{(4)}_{\bf 1}$.
Since the modular form that appears in the holomorphic modular-symmetric model also exists in the same form in the non-holomorphic modular-symmetric model, $Y^{(4)}_{\bf 1}$ depends only on $\tau$(or u). 
}
We can expand $Y^{(4)}_{\bf 1}$ as 
\begin{align}
Y^{(4)}_{\bf 1}(\tau) = -\frac{i}{\sqrt 3} \left. \frac{d\hat Y^{(4)}_{\bf 1}}{du} \right|_{u=0} (\tau -\omega)
+ \frac{8}{3}\left.\frac{d\hat Y^{(4)}_{\bf 1}}{du}\right|_{u=0} (\tau -\omega)^2+{\cal O}((\tau-\omega)^3).
\end{align}
Note that $Y^{(4)}_{\bf 1} (\tau = \omega)=0$.
For further details, see Appendix B.

Now, let us discuss the modulus stabilization 
by use of the above expansion of the modular forms.
Here, we use the unit $\Lambda=1$, where $\Lambda$ is a typical mass scale of the modulus stabilization.
We assume the following potential\footnote{{This term may be induced by non-perturbative effects. In supersymmetric modular flavor models, one assumes that the gaugino condensation leads to modular invariant modulus superpotential. 
Here, we discuss one scenario in non-supersymmetric models.
For example, we assume that the fermion $\Psi$ condensation leads to the modulus potential.
We start with the following Lagrangian:
$ K(\tau, \bar \tau)i\bar \Psi \gamma^\mu \partial_\mu \Psi + \frac{g(\tau,\bar \tau)}{M^2}(\bar \Psi \Psi)^2$.
We assume that $\Psi$ has the modular weight $-2$, i.e. $K(\tau, \bar \tau)=(2{\rm Im}\tau)^{-2}$. Then, the coefficient $g(\tau,\bar \tau)$ must be a real function with modular weights (4,4) for $\tau$ and $\bar \tau$ like $|Y^{(4)}_{\bf 1}(\tau)|^2$. Here, we use the canonically normalized field $\Psi'=\Psi/(2{\rm Im}\tau)$, which is modular invariant. Then, we assume their condensation, $\langle \bar \Psi' \Psi' \rangle \neq 0$. 
That leads to $\frac{\langle \bar \Psi \Psi \rangle ^2}{M^2} (2 {\rm Im}\tau)^{4} g(\tau,\bar \tau)$ as the modulus potential.}}
\begin{align}
 V(\tau,\bar \tau) = (2 {\rm Im}\tau)^{4} |Y^{(4)}_{\bf 1}(\tau)|^2.
 \label{eq:potential}
\end{align}
That is modular invariant.
It also satisfies the stationary condition:
\begin{align}
\frac{\partial V}{\partial \tau} = \frac{\partial V}{\partial \bar \tau} = 0,
\end{align}
at $\tau=\omega$.
In addition, we can calculate its second derivatives at $\tau=\omega$ as 
\begin{align}
\frac{\partial^2 V}{\partial \tau \partial \bar \tau} = 3 \left| \frac{d\hat Y^{(4)}_{\bf 1}}{du} \right|^2_{u=0}>0,  \qquad 
 \frac{\partial^2 V}{\partial \tau \partial  \tau} =\frac{\partial^2 V}{\partial \bar \tau \partial \bar \tau} = 0.
\end{align}
Thus, this potential has the minimum at $\tau = \omega$.

We have many varieties of modular invariant potential terms to choose.
For example, instead of $Y^{(4)}_{\bf 1} (\tau )$, we can use other moduli forms $Y^{(k)}_{\bf r}(\tau)$ in the potential (\ref{eq:potential}) with proper powers of $(2 {\rm Im}\tau)$.
If $Y^{(k)}_{\bf r}(\tau = \omega)=0$ and its first derivative does not vanish, the results are same and the modulus is stabilized at $\tau = \omega$.
However, if we choose the modular forms $Y^{(k)}_{\bf r}(\tau)$, which satisfies $Y^{(k)}_{\bf r}(\tau = \omega) \neq 0$ as well as non-vanishing first derivatives, and use the following potential:
\begin{align}
    v(\tau,\bar \tau) = (2 {\rm Im}\tau)^{2k} |Y^{(k)}_{\bf r}(\tau)|^2,
 \label{eq:potential-2}
\end{align}
the modulus is not be stabilized at $\tau = \omega$.
We need a proper choice of the potential term {to realize $\tau = \omega$. 
In general, the above potential can lead to other values of $\tau$. 
However, as discussed above, modular forms have remarkable properties around $\tau = \omega$.
Similarly, modular forms have specific properties around $\tau = i$.
Thus, the fixed points $\tau = \omega$ and $i$ as well as ${\rm Im} \tau=\infty$ are good candidates for stabilized values 
compared with other values.
}

Furthermore, the following combination of potential terms:
\begin{align}
     \tilde V(\tau,\bar \tau) &= V(\tau,\bar \tau) + \varepsilon v(\tau,\bar \tau) \notag \\
&=     (2 {\rm Im}\tau)^{4} |Y^{(4)}_{\bf 1}(\tau)|^2 + \varepsilon (2 {\rm Im}\tau)^{2k} |Y^{(k)}_{\bf r}(\tau)|^2,
 \label{eq:potential-3}
\end{align}
is also modular invariant.
Suppose $|\varepsilon| \ll 1$.
Then, the stationary condition $\partial \tilde V/\partial \tau =\partial \tilde V/\partial \bar \tau =0$ is approximated around $\tau = \omega$ as 
\begin{align}
    \frac{\partial^2 V}{\partial \tau \partial \bar \tau} \delta \bar \tau +\varepsilon \frac{\partial v}{\partial \tau} = 0, \qquad \frac{\partial^2 V}{\partial \tau \partial \bar \tau} \delta  \tau +\varepsilon \frac{\partial v}{\partial \bar \tau} = 0,
\end{align}
where $\delta \tau$ denotes a small deviation from $\tau = \omega$, i.e., 
$\tau = \omega + \delta \tau$.
Thus, by tuning $\varepsilon$, a small deviation from $\tau = \omega$ would be possible.

Another possibility for realizing a small deviation would be radiative corrections.
When we add proper effects, we may realize a small deviation.
For example, radiative corrections studied in ref.~\cite{Kobayashi:2023spx} can lead to 
a small deviation.\footnote{See also refs.~\cite{Higaki:2024jdk,Higaki:2024ueb} for radiative corrections.}
In the model, $A_4$ modular forms of the weight 8 with the representations ${\bf 1}'$ and ${\bf 1}''$ were used.
In addition, the model in ref.~\cite{Kobayashi:2023spx} corresponds to softly supersymmetry breaking model.
Thus, we replace the supersymmetric mass by the fermionic mass and the combination of supersymmetric mass squared and 
supersymmetric breaking mass squared by the bosonic mass squared.
Then, we can realize a small deviation from $\tau = \omega$ by choosing proper values of parameters.
In ref.~\cite{Ishiguro:2022pde}, a different type of corrections due to non-perturbative effects was studied.
Such corrections may lead to a small deviation from $\tau = \omega$.

\section{Conclusions and discussions}
\label{sec:conclusion}
We have studied the Zee-Babu model with a non-holomorphic modular $A_4$ symmetry, and
we have constructed the model so that there are minimum free parameters (two complex parameters for neutrino sector).
In this case, we have found only the normal hierarchy is allowed. 
%
%
Moreover, the allowed region to satisfy the neutrino oscillation data is localized at nearby $\tau=\omega$. The small absolute deviation, which is about 0.006, plays a crucial role in fitting these two mixings of $s^2_{23}$ and $s^2_{12}$.
In addition, we have obtained several predictions on Majorana and Dirac CP phases, and neutrinoless double beta decay as shown in our numerical analysis.  
{Also, we have discussed a possibility for the modulus stabilization within the framework of non-supersymmetric models.
Our modulus potential leads to the modulus stabilization at $\tau = \omega$.
A small deviation may be possible from $\tau = \omega$ by including proper corrections on the modulus potential.}
In the end, we have calculated the expansion of modular forms at nearby $\tau=\omega$ in the Appendix {B} so that one can apply them for a model and understand its analytical structure when one obtains the allowed region at nearby this fixed point.

\vspace{0.5cm}
\hspace{0.2cm} 

\begin{acknowledgments}
This work was supported in part JSPS KAKENHI Grant Numbers  JP23K03375 (T.K.).
H.O. is supported by Zhongyuan Talent (Talent Recruitment Series) Foreign Experts Project. 

\end{acknowledgments}

\appendix

\section{Concrete $A_4$ modular forms}
Here, we write down concrete $A_4$ modular forms in term of $q\equiv e^{2\pi i\tau}$ expansion~\cite{Qu:2024rns}.
$Y^{(0)}_3\equiv [y_1,y_2,y_3]^T$ is given by the following forms
\begin{align}
y_1& = y -\frac{3 e^{-4\pi y}}{\pi q}
 -\frac{9 e^{-8\pi y}}{2\pi q^2}
  -\frac{e^{-12\pi y}}{\pi q^3}
   -\frac{21 e^{-16\pi y}}{4\pi q^4}
    -\frac{18 e^{-20\pi y}}{5\pi q^5}
     -\frac{3 e^{-24\pi y}}{2\pi q^6} + \cdots\nn\\
     &-\frac{9\ln3}{4\pi} -\frac{3q}{\pi}
      -\frac{9q^2}{2\pi} -\frac{q^3}{\pi} -\frac{21q^4}{4\pi}
       -\frac{18q^5}{5\pi} -\frac{3q^6}{2\pi} +\cdots,\\
y_2& = \frac{27q^{1/3} e^{\pi y/3}}{\pi}
\left(\frac{e^{-3\pi y}}{4 q}
 +\frac{e^{-7\pi y}}{5 q^2}
  +\frac{5 e^{-11\pi y}}{16 q^3}
   +\frac{2 e^{-15\pi y}}{11 q^4}
    +\frac{2 e^{-19\pi y}}{7 q^5}
     +\frac{4 e^{-23\pi y}}{17 q^6} + \cdots\right)\nn\\
     &+\frac{9 q^{1/3}}{2\pi}\left(1+\frac{7q}{4}
      +\frac{8q^2}{7} +\frac{9q^3}{5} +\frac{14 q^4}{13}
       +\frac{31q^5}{16} + \frac{20 q^6}{19} +\cdots\right),\\
y_3& = \frac{9 q^{2/3} e^{2\pi y/3}}{2 \pi}
\left(\frac{e^{-2\pi y}}{q}
 +\frac{7e^{-6\pi y}}{4 q^2}
  +\frac{8 e^{-10\pi y}}{7 q^3}
   +\frac{9 e^{-14\pi y}}{5 q^4}
    +\frac{14 e^{-18\pi y}}{13 q^5}
     +\frac{31 e^{-22\pi y}}{16 q^6} + \cdots\right)\nn\\
     &+\frac{27 q^{2/3}}{\pi}\left(\frac14+\frac{q}{5}
      +\frac{5q^2}{16} +\frac{2 q^3}{11} +\frac{2 q^4}{7}
       +\frac{9 q^5}{17} + \frac{21 q^6}{20} +\cdots\right),
\end{align}
where $y\equiv {\rm Im}[\tau]$ and $Y^{(0)}_1=1$.
$Y^{(-2)}_3\equiv [y'_1,y'_2,y'_3]^T$ is given by the following forms
\begin{align}
y'_1& = \frac{y^3}3 +\frac{21 \Gamma_3(4\pi y)}{16\pi^3 q}
 +\frac{189 \Gamma_3(8\pi y)}{128\pi^3 q^2}
  +\frac{169 \Gamma_3(12\pi y)}{144\pi^3 q^3}
   +\frac{1533 \Gamma_3(16\pi y)}{1024\pi^3 q^4}
    + \cdots\nn\\
     &+\frac{\pi}{40}\frac{\zeta(3)}{\zeta(4)} 
     +\frac{21q}{8\pi^3}
      +\frac{189 q^2}{64\pi^3} +\frac{169 q^3}{72\pi^3} 
      +\frac{1533 q^4}{512\pi^3}
       +\frac{1323 q^5}{500\pi^3} +\frac{169 q^6}{64\pi^3} +\cdots,\\
y'_2& = -\frac{729q^{1/3}}{16\pi^3}
\left(\frac{\Gamma_3(8\pi y/3)}{16 q}
 +\frac{7\Gamma_3(20\pi y/3)}{125 q^2}
  +\frac{65\Gamma_3(32\pi y/3)}{1024 q^3}
   +\frac{74\Gamma_3(44\pi y/3)}{1331 q^4}
    + \cdots\right)\nn\\
     &-\frac{81 q^{1/3}}{16\pi^3}\left(1+\frac{73q}{64}
      +\frac{344q^2}{343} +\frac{567 q^3}{500} +\frac{20198 q^4}{2197}
       +\frac{4681 q^5}{4096}  +\cdots\right),\\
y'_3& = -\frac{81 q^{2/3}}{32\pi^3}
\left(\frac{\Gamma_3(4\pi y/3)}{q}
 +\frac{73\Gamma_3(16\pi y/3)}{64 q^2}
  +\frac{344\Gamma_3(28\pi y/3)}{343 q^3}
   +\frac{567\Gamma_3(40\pi y/3)}{500 q^4}
    + \cdots\right)\nn\\
     &-\frac{729 q^{2/3}}{8\pi^3}\left(\frac{1}{16}+\frac{7q}{125}
      +\frac{65 q^2}{1024} +\frac{74 q^3}{1331}   +\cdots\right),
\end{align}
where $\Gamma_3(z)\approx (z^2+2z+2)e^{-z}$.
The singlet with modular weight $-2$; $Y^{(-2)}_1$, is given by
\begin{align}
Y^{(-2)}_1& = \frac{y^3}3 -\frac{15 \Gamma_3(4\pi y)}{4\pi^3 q}
 -\frac{135 \Gamma_3(8\pi y)}{32\pi^3 q^2}
  -\frac{35 \Gamma_3(12\pi y)}{9\pi^3 q^3}
    + \cdots\nn\\
     &-\frac{\pi}{12}\frac{\zeta(3)}{\zeta(4)} 
     -\frac{15q}{2\pi^3}
      -\frac{135 q^2}{16\pi^3} -\frac{70 q^3}{9\pi^3} 
      -\frac{1095 q^4}{128\pi^3}
       -\frac{189 q^5}{25\pi^3} -\frac{35 q^6}{4\pi^3} +\cdots.
\end{align}
$Y^{(2)}_3\equiv [f_1,f_2,f_3]^T$ is given by the following forms~\cite{Feruglio:2017spp}:
\begin{align}
f_1& = 1 + 12 q+ 36 q^2+12 q^3+84q^4+72 q^5 +\cdots,\\
f_2& = -6 q^{1/3}
\left( 1 + 7 q+ 8 q^2+18 q^3+14q^4  +\cdots\right),\\
f_3& = -18q^{2/3}
\left( 1 + 2 q+ 5 q^2+4 q^3+8q^4 +\cdots\right).
\end{align}
Then, the modular forms with 4 modular weights can be obtained via $A_4$ multiplication rules~\cite{Ishimori:2010au} and their forms are as follows:
\begin{align}
Y^{(4)}_3&=[f^2_1-f_2f_3,f^2_3-f_1f_2,f^2_2-f_1f_3]^T,\\
Y^{(4)}_1&=f^2_1+2f_2f_3,\quad Y^{(4)}_{1'}=f^2_3+2f_1f_2,
\end{align}
where $Y^{(4)}_{1"}=0$.
Finally, the singlet modular forms with 8 weight is constructed by the ones of 4 modular weights as follows:
\begin{align}
Y^{(8)}_1=(f^2_1+2f_2f_3)^2,\quad
Y^{(8)}_{1'}=(f^2_1+2f_2f_3)(f^2_3+2f_1f_2),\quad Y^{(8)}_{1''}=(f^2_3+2f_1f_2)^2.
\end{align}

\section{$A_4$ modular forms and their derivatives}
In our paper, it is too difficult to {\it analytically} understand how the deviation from $\tau=\omega$ contributes to neutrino observables.
However, it would be worthwhile for us to discuss expansion modular Yukawa forms
in terms of the deviation.
We expect {\it in the future} that this forms would be helpful to understand analytical predictions for quark and lepton masses and mixings when one obtains solutions at nearby fixed points in framework of non-holomorphic modular symmetries.~\footnote{In whole the calculations in case of holomorphic case, refs.~\cite{Novichkov:2021evw,Kobayashi:2023spx,Abe:2024tox} are helpful.}

The modular symmetry transforms the modular form as
\begin{align}
 Y^{(k)}_{\bf r}(\gamma \tau, \overline{\gamma \tau}) = (c\tau + d)^k \rho_{\bf r}(\gamma) Y^{(k)}_{\bf r}(\tau, \bar \tau). 
\end{align}
In order to show the behavior of modular form around $\tau =\omega$, 
it is convenient to use the parameters $u$ and $\bar u$,
\begin{align}
u \equiv \frac{\tau -\omega}{\tau -\omega^2}, \ \bar u \equiv \frac{\bar \tau -\omega^2}{\bar \tau -\omega}.
\end{align}
The parameters $u$ and $\bar u$ transform to $\omega^2 u$ and $\omega \bar u$ by $ST$.
Also we define
\begin{align}
\hat Y^{(k)}_{\bf r}(u, \bar u) = (1-u)^{-k}Y^{(k)}_{\bf r}(\tau, \bar \tau).
\end{align}
Then, its transformation behavior under $ST$ is written as
\begin{align}
\hat Y^{(k)}_{\bf r}( \omega^2 u, \omega \bar u) = \omega^{-k} \rho_{\bf r}(ST) \hat Y^{(k)}_{\bf r}(u, \bar u).
\end{align}
When we focus on ${\bf r}={\bf 1}, {\bf 1}', {\bf 1}''$, this transformation behavior can be written 
by 
\begin{align}
\hat Y^{(k)}_{\bf r}( \omega^2 u, \omega \bar u) = \omega^{q_r-k} \hat Y^{(k)}_{\bf r}(u, \bar u),
\end{align}
where $q_r=0, 1, 2$ for ${\bf r}={\bf 1}, {\bf 1}',{\bf 1}''$, respectively.
We take the $\ell$-derivative of the above equation so as to obtain 
\begin{align}
  (\omega^{2\ell}- \omega^{q_r-k})\left.\frac{d^\ell \hat Y^{(k)}_{\bf r}}{du^\ell}\right|_{u=\bar u = 0} =0, 
\\
   (\omega^{\ell}- \omega^{q_r-k})\left.\frac{d^\ell \hat Y^{(k)}_{\bf r}}{d\bar u^\ell}\right|_{u=\bar u = 0} =0.
\end{align}

Next, we discuss ${\bf r}={\bf 3}$. 
The transformation behavior can be written by 
\begin{align}
\hat Y^{(k)}_{\bf 3}( \omega^2 u, \omega \bar u) = \omega^{-k}\rho_{\bf 3}(ST) \hat Y^{(k)}_{\bf 3}(u, \bar u), 
\end{align}
where 
\begin{eqnarray}
\rho_{\bf 3}(ST)=
\frac13
\left(
\begin{array}{ccc}
 -1 & 2 \omega & 2 \omega^2 \\
 2  & - \omega  & 2 \omega^2 \\
 2  & 2 \omega & - \omega^2 
\end{array}
\right).
\end{eqnarray}
The eigenvalues and the eigenvectors are 
\begin{eqnarray}
1 &:& v_1 = \frac13 (-1, 2 \omega, 2 \omega^2)^T, \\
\omega &:& v_\omega = \frac13 (2  , -\omega, 2 \omega^2)^T, \\
\omega^2 &:& v_{\omega^2} = \frac13 (2, 2\omega, -\omega^2)^T.
\end{eqnarray}
\begin{eqnarray}
\hat Y^{(k)}_{\bf 3} = \sum_{i = (1, \omega, \omega^2)} a_i^{(k)} v_i, 
\end{eqnarray}
where $v_i^\dagger v_j = \delta_{i j} $ and $a_i^{(k)} = v_i^\dagger \hat Y^{(k)}_{\bf 3} $.
We take the $\ell$-derivative of the above equation 
\begin{align}
  \omega^{2\ell}\left.\frac{d^\ell \hat Y^{(k)}_{\bf 3}}{du^\ell}\right|_{u=\bar u =0} &=&&
   \omega^{-k}\rho_{\bf 3}(ST)\left.\frac{d^\ell \hat Y^{(k)}_{\bf 3}}{du^\ell}\right|_{u=\bar u =0}
\\
  &=&& \omega^{-k}\rho_{\bf 3}(ST)\sum_{i = (1, \omega, \omega^2)} \left.\frac{d^\ell a_i^{(k)}}{du^\ell}\right|_{u=\bar u =0} v_i.
\end{align}
and 
\begin{align}
  \omega^{\ell}\left.\frac{d^\ell \hat Y^{(k)}_{\bf 3}}{d\bar u^\ell}\right|_{u=\bar u =0} &=&&
   \omega^{-k}\rho_{\bf 3}(ST)\left.\frac{d^\ell \hat Y^{(k)}_{\bf 3}}{d\bar u^\ell}\right|_{u=\bar u =0}
\\
  &=&& \omega^{-k}\rho_{\bf 3}(ST)\sum_{i = (1, \omega, \omega^2)} \left.\frac{d^\ell a_i^{(k)}}{d\bar u^\ell}\right|_{u=\bar u =0} v_i.
\end{align}
We can obtain the following equations, 
\begin{align}
  (\omega^{2\ell} - \omega^{q_i-k})\left.\frac{d^\ell a^{(k)}_i}{du^\ell}\right|_{u=\bar u =0} = 0, \ 
  (\omega^{\ell} - \omega^{q_i-k})\left.\frac{d^\ell a^{(k)}_i}{d\bar u^\ell}\right|_{u=\bar u =0} = 0. 
\end{align}

$k\equiv 0$ (mod 3) case

\begin{eqnarray}
\frac{d^\ell \hat Y^{(k)}_{\bf 3}}{du^\ell}
= \left\{
\begin{array}{ll}
\frac{d^\ell a_1^{(k)}}{d u^\ell} v_1 & (\ell \equiv 0 \  (mod \ 3))\\
\frac{d^\ell a_{\omega^2}^{(k)}}{d u^\ell} v_{\omega^2} & (\ell \equiv 1 \ (mod \ 3)) \\
\frac{d^\ell a_\omega^{(k)}}{d u^\ell} v_\omega & (\ell \equiv 2 \ (mod \ 3))
\end{array}
\right.
\end{eqnarray}
\begin{eqnarray}
\frac{d^\ell \hat Y^{(k)}_{\bf 3}}{d \bar u^\ell}
= \left\{
\begin{array}{ll}
\frac{d^\ell a_1^{(k)}}{d\bar u^\ell} v_1 & (\ell \equiv 0 \  (mod \ 3))\\
\frac{d^\ell a_{\omega}^{(k)}}{d\bar u^\ell} v_{\omega} & (\ell \equiv 1 \ (mod \ 3)) \\
\frac{d^\ell a_{\omega^2}^{(k)}}{d\bar u^\ell} v_{\omega^2} & (\ell \equiv 2 \ (mod \ 3))
\end{array}
\right.
\end{eqnarray}

$k\equiv 1$ (mod 3) case

\begin{eqnarray}
\frac{d^\ell \hat Y^{(k)}_{\bf 3}}{du^\ell}
= \left\{
\begin{array}{ll}
\frac{d^\ell a_\omega^{(k)}}{d u^\ell} v_\omega & (\ell \equiv 0 \  (mod \ 3))\\
\frac{d^\ell a_1^{(k)}}{d u^\ell} v_1 & (\ell \equiv 1 \ (mod \ 3)) \\
\frac{d^\ell a_{\omega^2}^{(k)}}{d u^\ell} v_{\omega^2} & (\ell \equiv 2 \ (mod \ 3))
\end{array}
\right.
\end{eqnarray}
\begin{eqnarray}
\frac{d^\ell \hat Y^{(k)}_{\bf 3}}{d \bar u^\ell}
= \left\{
\begin{array}{ll}
\frac{d^\ell a_\omega^{(k)}}{d\bar u^\ell} v_\omega & (\ell \equiv 0 \  (mod \ 3))\\
\frac{d^\ell a_{\omega^2}^{(k)}}{d\bar u^\ell} v_{\omega^2} & (\ell \equiv 1 \ (mod \ 3)) \\
\frac{d^\ell a_1^{(k)}}{d\bar u^\ell} v_1 & (\ell \equiv 2 \ (mod \ 3))
\end{array}
\right.
\end{eqnarray}

$k\equiv 2$ (mod 3) case

\begin{eqnarray}
\frac{d^\ell \hat Y^{(k)}_{\bf 3}}{du^\ell}
= \left\{
\begin{array}{ll}
\frac{d^\ell a_{\omega^2}^{(k)}}{d u^\ell} v_{\omega^2} & (\ell \equiv 0 \  (mod \ 3))\\
\frac{d^\ell a_{\omega}^{(k)}}{d u^\ell} v_{\omega} & (\ell \equiv 1 \ (mod \ 3)) \\
\frac{d^\ell a_1^{(k)}}{d u^\ell} v_1 & (\ell \equiv 2 \ (mod \ 3))
\end{array}
\right.
\end{eqnarray}
\begin{eqnarray}
\frac{d^\ell \hat Y^{(k)}_{\bf 3}}{d \bar u^\ell}
= \left\{
\begin{array}{ll}
\frac{d^\ell a_{\omega^2}^{(k)}}{d\bar u^\ell} v_{\omega^2} & (\ell \equiv 0 \  (mod \ 3))\\
\frac{d^\ell a_1^{(k)}}{d\bar u^\ell} v_1 & (\ell \equiv 1 \ (mod \ 3)) \\
\frac{d^\ell a_{\omega}^{(k)}}{d\bar u^\ell} v_{\omega} & (\ell \equiv 2 \ (mod \ 3))
\end{array}
\right.
\end{eqnarray}

\begin{align}
\frac{d u}{d \tau} = 
\frac{(u -1)^2}{i\sqrt{3}}, 
\ 
\frac{d \bar u}{d\bar \tau} = 
-\frac{(\bar u -1)^2}{i\sqrt{3}}. 
\end{align}

The first and second derivatives of the modular form $Y^{(k)}_{\bf r}$ are written by 
\begin{align}
& \frac{dY^{(k)}_{\bf r}}{d\tau}=\frac{(1-u)^{k+2}}{\sqrt 3 i}\left( \frac{d \hat Y^{(k)}_{\bf r}}{du}   - \frac{k}{1-u} \hat Y^{(k)}_{\bf r}\right), \\
& \frac{d^2Y^{(k)}_{\bf r}}{d\tau^2}=-\frac{(1-u)^{k+4}}{3}\left( \frac{d^2 \hat Y^{(k)}_{\bf r}}{du^2} -\frac{2(k+1)}{1-u} \frac{d \hat Y^{(k)}_{\bf r}}{du}   +\frac{k(k+1)}{(1-u)^2} \hat Y^{(k)}_{\bf r}\right), 
\end{align}
\begin{align}
& \frac{dY^{(k)}_{\bf r}}{d\bar \tau}=- \frac{(1-u)^{k} (1-\bar u)^2}{\sqrt 3 i}\frac{d \hat Y^{(k)}_{\bf r}}{d \bar u}  , \\
& \frac{d^2Y^{(k)}_{\bf r}}{d\bar \tau^2}=-\frac{(1-u)^{k} (1-\bar u)^4}{3}\left( \frac{d^2 \hat Y^{(k)}_{\bf r}}{d \bar u^2} - \frac{2}{1-\bar u}\frac{d \hat Y^{(k)}_{\bf r}}{du} \right), 
\\
& \frac{d^2 Y^{(k)}_{\bf r}}{d\tau d\bar \tau}= \frac{(1-u)^{k+2} (1-\bar u)^2}{3}
\left( \frac{d^2 \hat Y^{(k)}_{\bf r}}{d u d \bar u} - \frac{k}{1-u} \frac{d \hat Y^{(k)}_{\bf r}}{d \bar u}\right). 
\end{align}

By use of the above results, we can expand the modular form around $\tau = \omega$ and $\bar \tau = \omega^2$.
We can expand $Y^{(k)}_{\bf r}$ as 
\begin{eqnarray}
 Y^{(k)}_{\bf r}(\tau, \bar\tau) 
 &=& 
\hat Y^{(k)}_{\bf r}(0, 0)  
 - \frac{i}{\sqrt 3}\left( \frac{d \hat Y^{(k)}_{\bf r}(0, 0)}{d u}- k \hat Y^{(k)}_{\bf r}(0, 0) \right)  (\tau - \omega) 
\nonumber\\
 &&
+ \frac{i}{\sqrt 3}\frac{d \hat Y^{(k)}_{\bf r}(0, 0)}{d \bar u} (\bar \tau - \omega^2)  + {\cal O}(|\tau - \omega|^2). 
\end{eqnarray}

{Applying the above discussion,
we show more useful forms of $Y^{(0)}_3$ and $Y^{(-2)}_3$ at nearby $\tau=\omega$ as follows:
\begin{eqnarray}
 Y^{(0)}_{\bf 3}(\tau, \bar\tau) 
&=& b^{(0)}_0 v_1 - \frac{ i b^{(0)}_1}{\sqrt 3}(\epsilon v_{\omega^2} - \bar \epsilon v_{\omega})   + {\cal O}(|\epsilon|^2), 
\\
 Y^{(-2)}_{\bf r}(\tau, \bar\tau) 
  &=& b_0^{(-2)} v_{\omega}
 - \frac{i}{\sqrt 3}\left( b_1^{(-2)} v_{1}+2 b_0^{(-2)} v_{\omega} \right)  \epsilon 
 + \frac{i}{\sqrt 3} b_2^{(-2)} v_{\omega^2} \bar \epsilon 
   + {\cal O}(|\epsilon|^2).
\end{eqnarray}
The definitions of $b^{(k)}_i$ and values are followings: 
\begin{eqnarray}
\hat {\bf Y}_{\bf 3}^{(0)}(0, 0)
=
b^{(0)}_0 v_{1}, 
\frac{d\hat {\bf Y}_{\bf 3}^{(0)}}{du}(0, 0)
=
b^{(0)}_1  v_{\omega^2}, 
\frac{d \hat {\bf Y}_{\bf 3}^{(0)}}{d\bar u}(0, 0)
=
b^{(2)}_2  v_{\omega}, 
\\
\hat {\bf Y}_{\bf 3}^{(-2)}(0, 0)
=
b^{(-2)}_0 v_{\omega}, 
\frac{d\hat {\bf Y}_{\bf 3}^{(-2)}}{du}(0, 0)
=
 b^{(-2)}_1  v_{1}, 
\frac{d \hat {\bf Y}_{\bf 3}^{(-2)}}{d\bar u}(0, 0)
=
b^{(-2)}_2  v_{\omega^2}, 
\end{eqnarray}
where 
$b^{(0)}_0=-0.2623\equiv b_0$, $b^{(0)}_1=b^{(0)}_2 =1.232, b^{(-2)}_0=0.4166, b^{(-2)}_1=-0.1543, b^{(-2)}_2 =1.315$.

Then, we can diagonalize the mass matrix for the charged-lepton as the leading order, 
\begin{align}
M_\ell^0  = \frac{v}{\sqrt2}
\left[
a_e \begin{pmatrix}
1 & 0 & 0 \\ 
 0 & 0 &1 \\ 
0 & 1 & 0 \\ 
\end{pmatrix}
+ 
 \frac{b_e b^{(0)}_0}{3} 
\begin{pmatrix}
-2& -2 \omega^2& -2 \omega\\
-2 \omega^2& 4 \omega& 1\\
-2 \omega& 1 &4 \omega^2\\
\end{pmatrix}
+ 
 \frac{c_e b^{(0)}_0}{3} 
\begin{pmatrix}
0 & 2 \omega^2& -2 \omega\\
-2 \omega^2& 0 & -1\\ 2 \omega& 1& 0\\
\end{pmatrix}
\right]
 \label{massmat_0},
\end{align}
and $V_{e_L}^0$ and $V_{e_R}^0$ are given by
\begin{align}
V_{e_L}^0  =  
\begin{pmatrix}
1 & 0 & 0 \\ 
 0 & \omega^* & 0 \\ 
0 & 0 & \omega \\ 
\end{pmatrix}
\begin{pmatrix}
2/3 & 2/3 & -1/3 \\ 
-1/3 & 2/3 & 2/3 \\ 
2/3 & -1/3& 2/3 \\ 
\end{pmatrix},\\
V_{e_R}^0  =  
\begin{pmatrix}
1 & 0 & 0 \\ 
 0 & \omega & 0 \\ 
0 & 0 & \omega^* \\ 
\end{pmatrix}
\begin{pmatrix}
-2/3 & 2/3 & -1/3 \\ 
-2/3 & -1/3 & 2/3 \\ 
1/3 & 2/3 & 2/3 \\ 
\end{pmatrix}
 \label{chgd-mix_0},
\end{align}
where $a_e=0.00226178$, $b_e=0.00944476$, $c_e=0.000742539$ are inserted that are uniquely determined by three charged-lepton mass eigenvalues.

}

\section{Confidence level}

We discuss the calculation of the CL. 

The probability density function is written by the following form: 
\begin{eqnarray}
 f(x, \nu)= \frac{x^{\nu/2-1} \exp (\frac{x}{2})}{2^{\nu/2} \Gamma (\frac{\nu}{2})},  
\end{eqnarray}
where $\nu$ is the degree of freedom (DOF). 
It is normalized by 
\begin{eqnarray}
 \int^\infty_0 f(x, \nu) dx = 1. 
\end{eqnarray}

The CL for $\nu$ DOF is given by 
\begin{eqnarray}
 \int_0^{\Delta \chi^2} f(x, \nu) dx,  
\end{eqnarray}
where $\Delta \chi^2 = \sum_{i=1}^{\nu} \left( \chi_i^2 - \chi^2_{i, min} \right)$ 
and all parameters are independent.
We can get the values $\chi_i^2$ and $\chi_{i, min}^2$ from NuFit6.0~\cite{Esteban:2024eli} . 

\begin{table}[thb]
  \begin{tabular}{|c||c|c|c|c|c|c|}
  \hline
    CL(\%)                & $\nu =1$ & $\nu =2$ & $\nu =3$ & $\nu =4$ & $\nu =5$ & $\nu =6$ \\
  \hline
    68.27 (1$\sigma$)  & 1.00		  & 2.30		& 3.53		   & 4.72		& 5.89	    	  &	 7.04		\\
  \hline
    95.45 (2$\sigma$)  & 4.00		  & 6.18		& 8.02		   & 9.72		& 11.31     &	 12.85		\\
  \hline
    99.73 (3$\sigma$)  & 9.00		  & 11.83		& 14.16	   & 16.25		& 18.21   	  &	 20.06		\\
  \hline
    100$-$5.7$\times$10$^{-5}$ (5$\sigma$) 
     & 25.00		  & 28.74		& 31.81	   & 34.56		& 37.09   	  &	 39.49		\\
  \hline
\end{tabular}
\caption{Values $\Delta \chi^2$ corresponding to CL for joint estimation of $\nu$ parameters.}
\label{tab:chiCL}
\end{table}
Tab.~\ref{tab:chiCL} shows values $\Delta \chi^2$ corresponding to CL for joint estimation of $\nu$ parameters.

{
\section{The neutrino mass matrix at $\tau=\omega$}
Modular forms at the fixed point are the followings,
\begin{eqnarray}
{\bf Y}_{\bf 3}^{(0)}
&=&
b_0 v_{1}, 
\\
{\bf Y}_{\bf 3}^{(-2)}
&=&
b_{-2} v_{\omega}, 
\end{eqnarray}
where $b_{0}=-0.2623$, $b_{-2}=0.4166$.

At the fixed point, $\tilde{b}_k = 0$ and the neutrino mass matrix is given by 
\begin{eqnarray}
m_\nu = \frac{\kappa v^2}{9} b_{0}^2 b_{-2}(a_e - b_0 (b_e+c_e))^2
\begin{pmatrix}
4 & - 2 \omega^2 & 4 \omega \\ 
-2 \omega^2 & \omega & -2 \\ 
4\omega & -2 & 4\omega^2 \\ 
\end{pmatrix}. 
\end{eqnarray}
The mass matrix is rank 1 and has two massless states. 

}

\bibliography{nhm4zb.bib}

\end{document}